\documentclass[%
 reprint,
 amsmath,amssymb,
 aps,
]{revtex4-2}

\usepackage{graphicx}
\usepackage{dcolumn}
\usepackage{bm}

\usepackage{mathptmx}
\usepackage{url}
\usepackage{subcaption} 
\usepackage{multirow}

\begin{document}

\preprint{APS/123-QED}

\title{Wave Packet Propagation in Tilted Weyl Semimetals for Black Hole Analog Systems}

\author{M.\ A.\ Lozande}
 \affiliation{Department of Physics, Mindanao State University - Main Campus, Marawi City\\Lanao del Sur, 9700, Philippines}
\author{E.\ A.\ Fajardo}%
\affiliation{Department of Physics, Mindanao State University - Main Campus, Marawi City\\Lanao del Sur, 9700, Philippines}%

\date{\today}

\begin{abstract}
We explore the realization of distinct analog black hole horizons within tilted Weyl semimetals by comparing two models with contrasting spectral properties. We demonstrate that a spatially varying tilt in the Weyl cone structure creates an effect analogous to the tilting of light cones near a gravitational black hole horizon. By analyzing wave packet dynamics in both models, we reveal two fundamentally different types of analog horizons. The first model exhibits complete wave packet reflection, effectively mimicking an impenetrable barrier. In contrast, the second model permits wave packet transmission across the horizon. Critically, for both models, wave packets initialized with zero momentum ($k_0=0.0$) experience the strongest horizon effects, characterized by a dramatic slowing and significantly longer dwell times at the horizon region. Finally, we find that both systems exhibit substantial probability loss, which we demonstrate is directly correlated with the wave packet's dwell time near the horizon. Our findings establish tilted Weyl semimetals as a rich, tunable platform for investigating non-trivial quantum effects and information dynamics associated with analog black hole horizons.
\end{abstract}

\maketitle

\section{\label{sec:level1}Introduction}

\begin{figure}
  \centering{\includegraphics[width=0.8\linewidth] {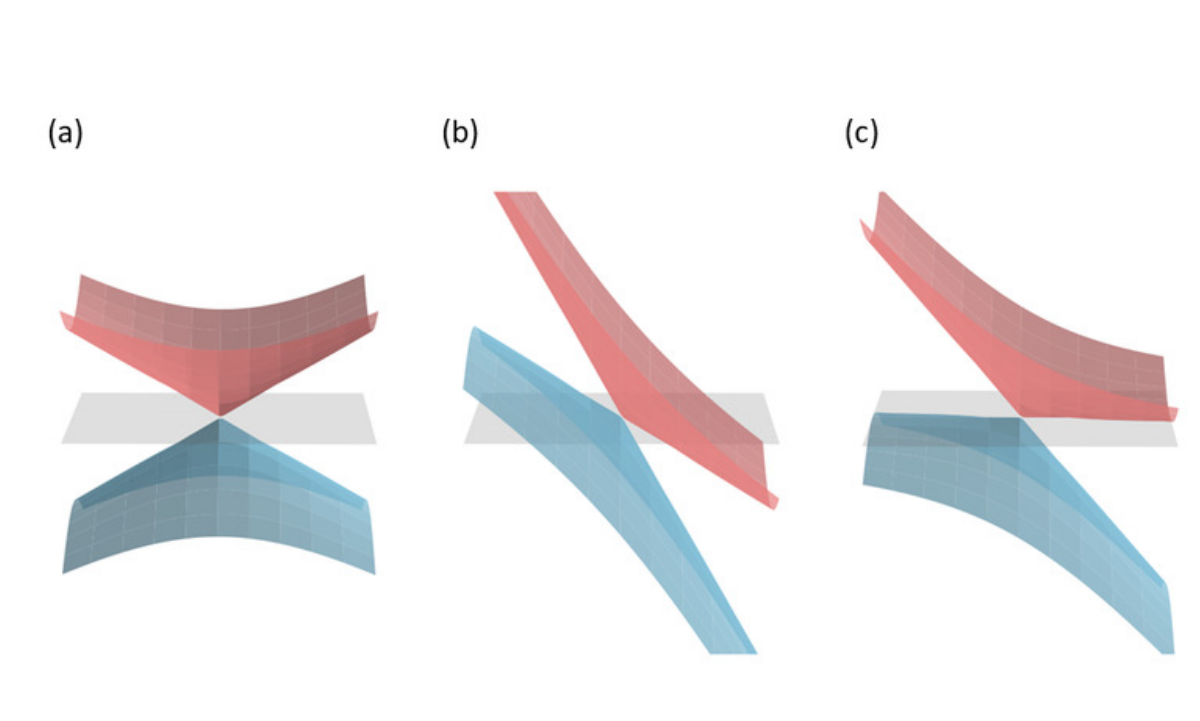}}
  \caption{Types of Weyl semimetals: (a) untilted type-I with point-like Fermi surface, (b) strongly tilted type-II with electron and hole pockets and (c) critically tilted type-III Weyl cone.}\label{fig:weyl-type}
\end{figure}

Weyl semimetals (WSMs) are topological quantum materials whose low-energy electronic excitations near the Weyl nodes behave like the massless relativistic particles described by the Weyl equation \cite{weyl1929}. Since their theoretical prediction, WSMs have been experimentally realized in a growing number of materials, including TaAs \cite{lv2015a,lv2015b,xu2015a,yang2015}, NbAs \cite{xu2015b}, TaP \cite{xu2016}, and NbP \cite{shekhar2015}. The intriguing feature of WSMs is that near the Weyl nodes where the valence and conduction bands touch linearly, the low-energy excitations of the electrons resemble the elementary particles proposed in high-energy physics. This linearity is protected by the topological properties of the band structure, but it's not a fundamental property of the constituent particles. At higher energies, away from the Weyl points, the dispersion relation deviates from linearity due to the lattice effects. The periodic potential of the lattice introduces higher-order terms in the energy-momentum relation, causing the dispersion to become nonlinear and the Weyl fermions no longer behave as strictly massless particles. See Refs.~\cite{armitage2018} and \cite{rao2016} for more details. What makes WSMs particularly interesting in the context of this work is that a spatially varying tilt of the Weyl cone is mathematically analogous to the tilting of the light cone in general relativity \cite{haller2023,konye2022,konye2023,sabsovich2022}. WSMs are classified as type-I, type-II \cite{soluyanov2015}, or type-III \cite{huang2018} depending on the strength of the tilt as shown in Fig.~\ref{fig:weyl-type}. Interestingly, the tilting of the Weyl cones create regions where the negative energy states of the original positive branch become empty while the positive energy states of the original negative branch become occupied. This process corresponds to the creation of electron-hole pairs, with one partner escaping as radiation and the other being trapped analogous to the physics of Hawking radiation \cite{sabsovich2022,huang2018,volovik2016,de2021}.

In this work, we focus on wave packet dynamics within a fixed spacetime geometry, providing a foundational analysis of the system's behavior. We first discuss, in Sec.~\ref{sec:tilted-weyl-models}, the simplified one-dimensional models exhibiting Weyl-like behavior, which serve as the foundation for constructing Weyl cones and simulating their dynamics under a hyperbolic tangent tilt profile. In Sec.~\ref{sec:energy-disp-metric} we establish the correspondence between the dispersion relation for the Weyl Hamiltonian with spatially varying tilt to the Weyl fermions in curved spacetime for a metric $g^{\mu \nu}$. For simplicity, we use the Schwarzschild metric for the black hole written in Painleve-Gullstrand form. Section~\ref{sec:wave-packet-prop} presents the time evolution of the wave packet obtained using the fourth-order Runge-Kutta (RK4) method and cross-validated with Wentzel-Kramers-Brillouin (WKB) approximation. Finally, we analyze the scattering coefficients and the dwell time of the wave packet at the horizon.

\section{\label{sec:tilted-weyl-models}Models for Tilted Weyl Cone}

In condensed matter physics, the low energy Hamiltonian captures the essential features of a system's low-energy excitations around the Fermi energy. For a Weyl continuum Hamiltonian with a tilted Weyl cone, the equation is given by,
\begin{equation}
  \label{eq:weyl-ham-tilt}
  H=\mathbf{\sigma} \cdot \mathbf{k} - \sigma_0 (\mathbf{V} \cdot \mathbf{k}),
\end{equation}
where $\mathbf{k}$ is the momentum, $\mathbf{V}$ is the tilt of the system, $\sigma_0$ denotes the identity matrix and the components $\sigma_i$ are the standard $2 \times 2$ Pauli matrices. By diagonalizing the Hamiltonian in Eq.~(\ref{eq:weyl-ham-tilt}), the energy dispersion relation is
\begin{equation}
  \label{eq:disp-ham-tilt}
  E_{\pm} = \pm |\mathbf{k}| - \mathbf{V} \cdot \mathbf{k}.
\end{equation}
Here the plus and minus signs corresponds to the positive and negative energy states of the band structure respectively. Since the tilt term is proportional to the identity matrix, it does not affect the eigenstates which means that a wave packet initially at $E_+$ ($E_-$) will remain in the positive energy branch (negative energy branch) as it evolves in time. Note that while the nodal tilt $V$ in Eq.~(\ref{eq:disp-ham-tilt}) describes a Weyl node with constant tilt, we focus on spatially varying tilts to study wave packet propagation.

\begin{figure}[bt]
  \centering{\includegraphics[width=0.8\linewidth] {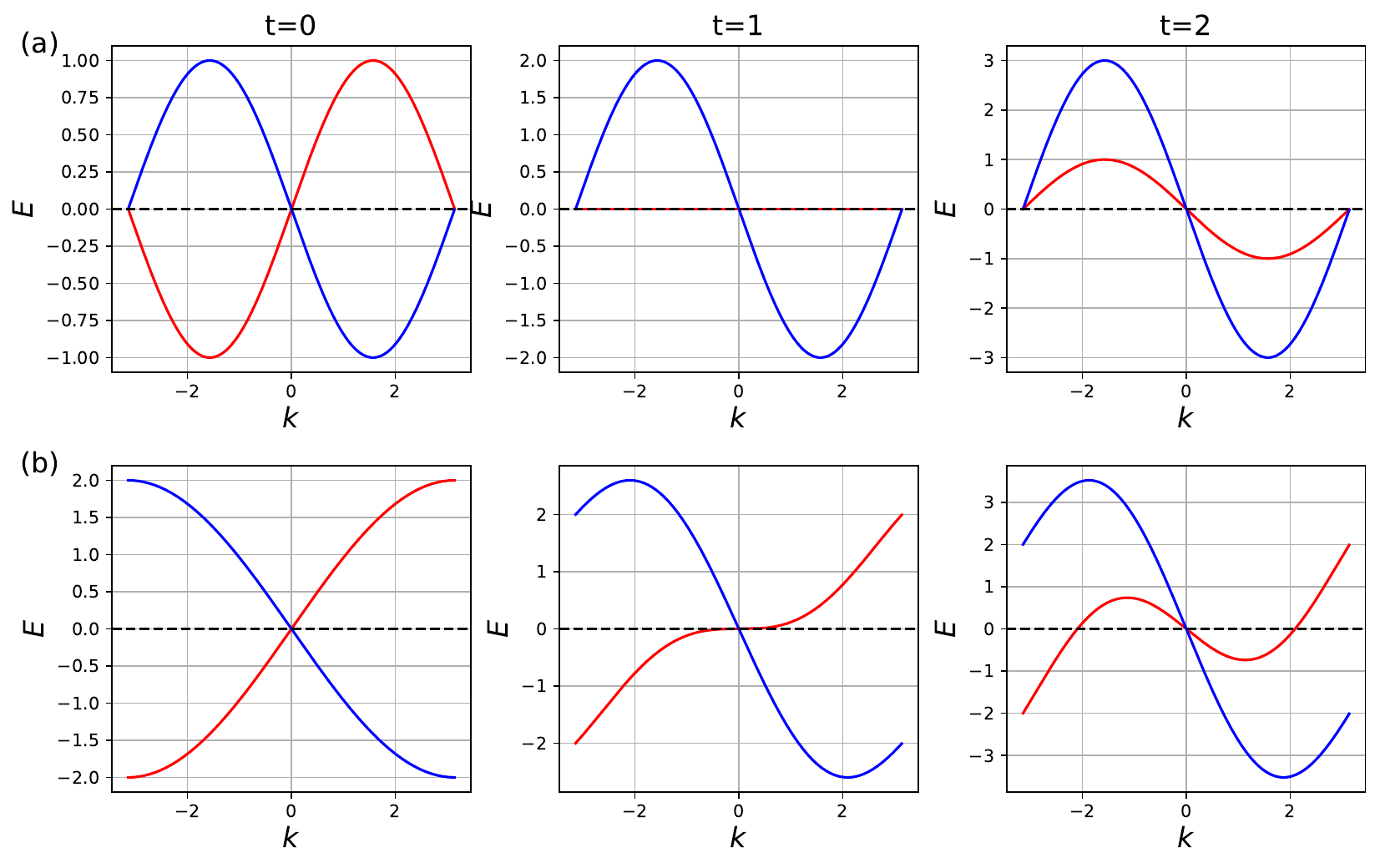}}
  \vspace*{8pt}
  \caption{Energy dispersion plot of (a) $E_1$ and (b) $E_2$ with different tilt values. The red (blue) band corresponds to the positive (negative) energy branch. Fermi energy $E_F$ is highlighted by the black broken lines.}\label{fig:energy-disp}
\end{figure}

For simplicity, we focus on the $x$ direction and set $\textbf{k} \equiv k_x$. The low-energy Hamiltonian described in Eq.~(\ref{eq:weyl-ham-tilt}) can then be consequently mapped to two lattice models
\begin{equation}
\label{eq:Weyl_Hamiltonian_1}
H_1 = (\sigma_x - V \sigma_0) \sin k,
\end{equation}
and
\begin{equation}
\label{eq:Weyl_Hamiltonian_2}
H_2 = (\sigma_x - V \sigma_0) \sin k + \sigma_z (1 - \cos k),
\end{equation}
which behave similarly at low $k$ values but differ at $k>0$. Using Taylor expansion about $k=0$, these two models reproduce Eq.~(\ref{eq:weyl-ham-tilt}). The respective energy dispersion relations for $H_1$ and $H_2$ are
\begin{equation}
  \label{eq:energy-disp}
E_{1\pm} = \pm \sin k - V \sin k,
\end{equation}
and
\begin{equation}
\label{eq:energy-disp2}
E_{2\pm} = \pm \sqrt{\sin^2 k + (1 - \cos k)^2} - V \sin k.
\end{equation}

Near the Weyl nodes, the dispersion is linear and implies that the fermions behave like massless particles. However, the linear energy dispersion characteristic of Weyl fermions is not indefinite. As the momentum $k$ moves further away from the Weyl node within the periodic Brillouin zone, the higher order terms in the band structure become significant which breaks the linearity. Notice that in the first model (shown on Fig.~\ref{fig:energy-disp}(a)), there is a second Weyl node at the edge of the Brillouin zone whereas the second model (Fig.~\ref{fig:energy-disp}(b)) has only one Weyl node. At $V=0$, both models are symmetric but at $V=2$ the Weyl cone is overtilted creating electron and hole pockets which simulates the interior of a black hole discussed in Refs.~\onlinecite{volovik2016} and \onlinecite{zubkov2018}. 

Between the interface of type-I and type-II WSMs is the position of the analog event horizon. Using $V=1$, one can see that the eigenenergies (red band) on the first model are zero for all values of $k$ causing the group velocity in this region to be zero as well. On the other hand, one can see that there are non-zero energy states for $k>0$ when we use the second model.

In addition to the above mentioned difference on both models, the electron-hole pocket extends throughout the entire Brillouin zone on $E_1$ but is finite and localized on $E_2$. Interestingly, the red band of the overtilted type-II Weyl point of $E_1$ does not cross the zero-energy Fermi level if we go from $k=0$ to $k=3$ but it does on the latter model. These differences between the two models will have significant effects on the dynamics of our wave packet as also shown in Ref.~\onlinecite{konye2022}.

\subsection{Event horizon in tanh tilt profile}

Recent studies have demonstrated the tunability of Weyl nodes in real materials can be dynamically tuned, leading to transitions between type-I and type-II WSM. For instance, a study on $\mathrm{MnSb_2Te_4}$ showed that external perturbations can drive the system from a type-I towards a type-II WSM behavior \cite{tamanna2024}. Conversely, theoretical work on $\mathrm{MoTe_2}$ predicts that strain or correlation effects could induce a type-II to type-I WSM transition \cite{sun2015}. Furthermore, Ref.~\onlinecite{kong2017} proposed a tunable two-band lattice model in an ultracold atom setup where the tilt of the Weyl nodes can be controlled through the Hamiltonian parameters, enabling simulations of type-I, type-II, and even a hybrid type-1.5 WSM.

To model a realistic analog black hole geometry with a smooth event horizon, we implement a position-dependent tilt profile using a hyperbolic tangent function:
\begin{equation}
V(x) = V_{\text{max}} \cdot \frac{\tanh\left(\frac{x - x_h}{a}\right) - \epsilon}{1 - \epsilon}
\end{equation}
where $V_{\text{max}} = 2$ is the maximum tilt value, $x_h$ is the horizon position, $a$ is the transition width parameter controlling the spatial scale over which the tilt of the Weyl cone varies appreciably, and $\epsilon = \tanh\left(\frac{x_0 - x_h}{a}\right)$ is the normalization factor with $x_0$ representing the initial wave packet position.

This normalization ensures that the tilt profile begins at zero at the initial wave packet position $x_0$, representing an untilted region where the wave packet is initialized, while the denominator ensures the maximum tilt reaches exactly $V_{\text{max}}$ at large $x$. The profile creates a smooth transition from a type-I WSM ($V < 1$) to a type-II WSM ($V > 1$), with the analog event horizon located at the point where $V(x) = 1$. 

In the simulation, the analog event horizon naturally appears when $V=1$ which occurs at $x\approx 1884.5$ for our parameter choices providing sufficient space on both sides of the horizon for clear observation of wave packet dynamics. As mentioned above, the parameter $a$ in our model can be interpreted as a spatial scale over which the tilt of the Weyl cone varies appreciably. While $a$ is not directly measurable in experiments, it serves as a phenomenological parameter that captures the smoothness of tilt transitions. We set $a = 700$ to ensure the tilt varies slowly compared to the width of the wave packet. This choice  is motivated by the desire to maintain approximate adiabaticity in the wave packet propagation. Further work may investigate the sensitivity of our results to different values of $a$, potentially guided by specific experimental realizations in WSMs.

\section{\label{sec:energy-disp-metric}Energy Dispersion from the Metric}

In the following, we explore the relation between the tilted WSM  and black holes. Importantly, we show the equivalence of the dispersion relation for a metric $g^{\mu \nu}$ following the sign convention (--+++) to the dispersion of a Weyl system. 

The untilted Weyl cone corresponds to a light cone in flat spacetime while the tilted and overtilted Weyl cone corresponds to a light cone that moves closer to a black hole  \cite{sabsovich2022}. Using these analogies, we can map the tilted Weyl Hamiltonian in Eq.~(\ref{eq:weyl-ham-tilt}) to the spacetime metrics where the motion of massless particles is described by the null geodesics
\begin{equation}
ds^2 = g_{\mu \nu} dx^{\mu} dx^{\nu} = 0.
\end{equation}
For massless Weyl fermions, the dispersion relation follows from the Dirac equation in curved spacetime
\begin{equation}
  \label{eq:dirac-spacetime}
g^{\mu \nu} p_{\mu} p_{\nu}= 0,
\end{equation}
where $g^{\mu \nu}$ is the contravariant form of the metric tensor and $p_{\mu} = (E,k)$ is the energy-momentum four-vector. The spacetime metric relates to black hole using Einstein's field equation \cite{einstein1915} and for our purpose, the Schwarzschild metric (see Ref.~\onlinecite{blinn2017} for detailed derivation) written in Painlev\'{e}-Gullstrand form which satisfies
\begin{equation}
  \label{eq:PG-schwarzschild-metric}
  ds^2=-(1 - V^2)dt^2 - 2tdxdt + dx^2,
\end{equation}
works well. The $V^2 - 1=0$ term can be considered as the tilt and in this sense, an event horizon exists at $V=1$ \cite{kedem2020}. In matrix form, Eq.~(\ref{eq:PG-schwarzschild-metric}) can be written as
\begin{equation}
g^{\mu \nu} = \begin{pmatrix} -1 & \quad -V  \\ -V & \quad  1-V^2 \end{pmatrix},
\end{equation}
and so we can rewrite Eq.~(\ref{eq:dirac-spacetime}) as
\begin{equation}
  \label{eq:dirac-disp}
  -E^2 - 2VEk + (1 - V^2)k^2=0.
\end{equation}
The two solutions for Eq.~(\ref{eq:dirac-disp}) gives the dispersion relation of a tilted Weyl cone
\begin{equation}
E_{\pm} = \pm k - Vk,
\end{equation}
which is physically equivalent to Eq.~(\ref{eq:disp-ham-tilt}) derived from the Weyl Hamiltonian.

\section{\label{sec:wave-packet-prop}Wave Packet Propagation}

To simulate wave packet propagation, we use a normalized Gaussian packet using
\begin{equation}
\psi(x,0) = \mathcal{N} \exp\left(-\frac{(x-x_0)^2}{2\sigma^2}\right) \exp(ik_0(x-x_0))
\label{eq:initial_wavefunction}
\end{equation}
where $\mathcal{N}$ is the normalization constant, $\sigma = 20$ is the wave packet width, and $k_0 \in \{0.0, 0.1, 0.2\}$ represents the different initial momentum values that corresponds to the near Weyl node relativistic regime. The spatial domain extends over $L = 4000$ grid points with uniform spacing $\Delta x = 1$. We tracked the center of the wave packet using
\begin{equation}
    \langle x \rangle (t) = \frac{\int x|\psi(x,t)|^2 dx}{\int |\psi(x,t)|^2dx}
\end{equation}
which represents the expectation value of position at time $t$.

In this work, periodic boundary conditions are implemented to ensure well-defined spatial derivatives at all grid points and to prevent artificial boundary reflections. However, the simulation domain is chosen sufficiently large such that the wave packet and its scattered components remain well-separated from the boundaries for the entire simulation duration. As a result, the wave packet never interacts with the boundaries, and the periodic conditions remain inactive. This design choice prevents artificial wrap-around effects and ensures that the computed transmission and reflection properties are free from boundary artifacts.

\begin{figure}
    \centering
    \includegraphics[width=1.0\linewidth]{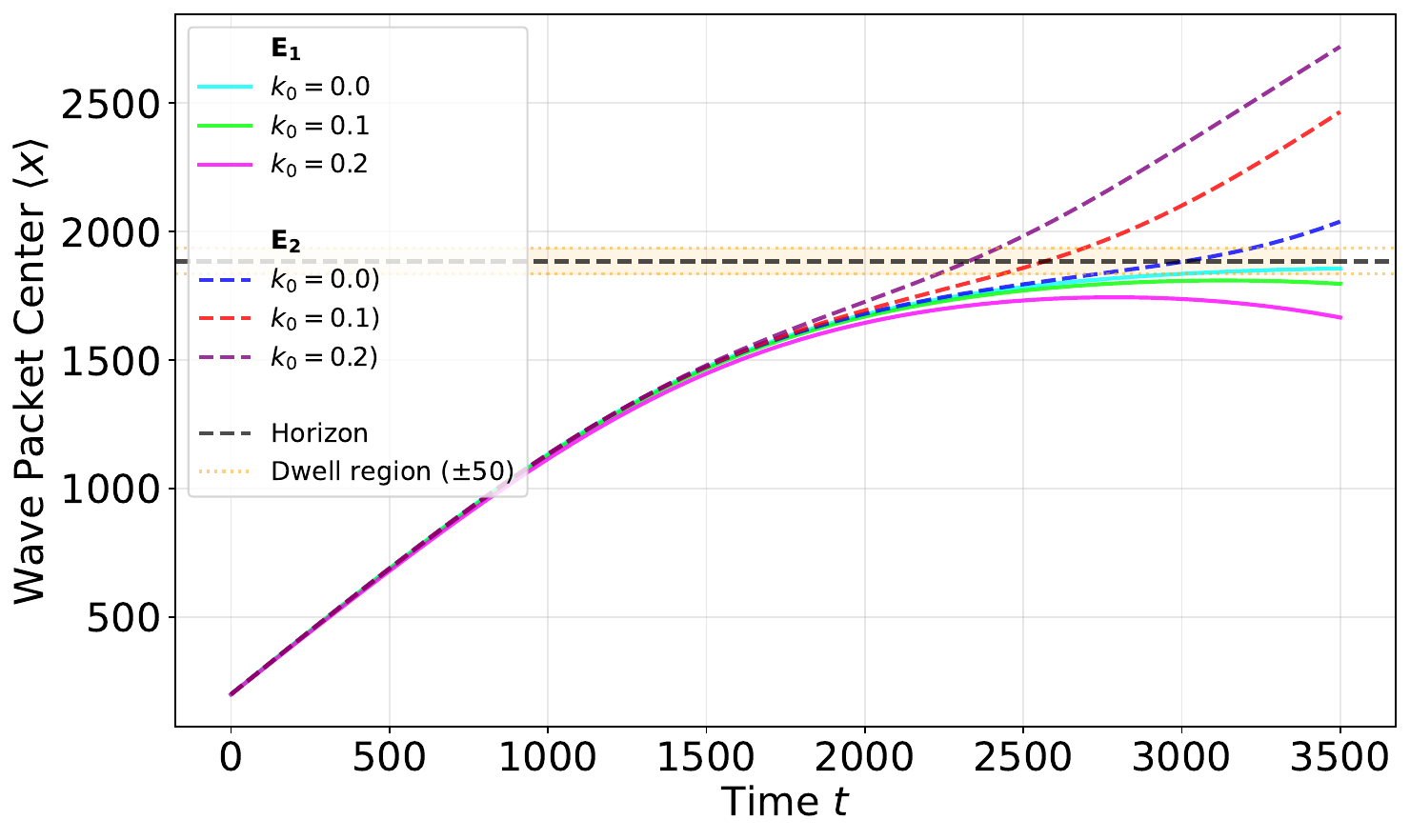}
    \caption{Wave packet center of mass $\langle x \rangle$ vs. time $(t)$ for $E_1$ (straight lines) and $E_2$ (broken lines) with initial momenta $k_0 = 0.0, 0.1, 0.2$. The shaded region marks the horizon dwell zone ($x_h \pm 50$). First model $E_1$ shows complete reflection with momentum-dependent penetration depths while second model $E_2$ demonstrates successful transmission for all packets with decreasing dwell times at higher momenta.}
    \label{fig:wave_trajectories}
\end{figure}

To study the dynamics of Weyl fermions in curved spacetime analogs, we employ multiple complementary approaches. The wave packet evolution is computed using two methods: a numerical simulation based on a fourth-order Runge–Kutta (RK4) scheme and a semiclassical analysis using the Wentzel–Kramers–Brillouin (WKB) approximation. To quantify the interaction with the analog event horizon, we extract scattering observables from the spatial distribution of the final wave packet and characterize the temporal dynamics through dwell time measurements. Together, they provide a consistent and comprehensive description of wave packet propagation near the analog event horizon.

\subsection{\label{sec:RK4}Runge–Kutta method}
In the long-wavelength limit ($k \ll 1$), we can approximate $\sin(k) \approx k$ using Taylor series expansion. Thus, the energy dispersion from Eq.~(\ref{eq:energy-disp}) in real space becomes
\begin{equation}
\label{eq:linear_dispersion}
    E_1 \approx \pm(1 - V) k.
\end{equation}

In our simulations we considered only the positive branch of the dispersion relation. The two branches correspond to opposite group velocities, so the negative branch produces dynamics that are identical in form but mirror the positive branch in direction. Since the focus of this work is on the qualitative behavior of wave packet propagation rather than on the symmetry between opposite directions, restricting to the positive branch is sufficient and avoids redundancy in the analysis.

To describe the wave packet dynamics, we start by using the dispersion relation from Eq.~(\ref{eq:linear_dispersion}) by applying the standard correspondence $k \rightarrow -i \hbar \partial_x$. From this, we obtain the effective Hamiltonian operator for the first model
\begin{equation}
\label{eq:hamiltonian1_op}
\hat{H}_1 = (1 - V)(-i\hbar\partial_x).
\end{equation}
This operator is then used in the time-dependent Schrödinger equation (TDSE)
\begin{equation}
i\hbar \partial_t \psi (x,t) =\hat{H} \psi(x,t).
\end{equation}
Substituting the Hamiltonian from Eq.~(\ref{eq:hamiltonian1_op}) and dividing both sides by $i$ gives
\begin{equation}
\label{eq:TDSE_1}
\partial_t \psi = -\alpha \partial_x \psi
\end{equation}
where $\alpha = 1-V$ and $\hbar =1$.

To discretize the spatial derivatives in Eq.~(\ref{eq:TDSE_1}), we employ the central difference scheme by approximating the derivatives on a discrete grid
\begin{equation}
\label{eq:spatial_derivative}
\partial_x \psi \approx \frac{\psi_{i+1} - \psi_{i-1}} {2 \Delta x},
\end{equation}
where $\psi_{i+1}$ ($\psi_{i-1}$) is the value of the wave function at the next (previous) grid point and $\Delta x$ is the spatial step size between the adjacent grid points. Plugging in Eq.~(\ref{eq:spatial_derivative}) into the right-hand side of Eq.~(\ref{eq:TDSE_1}), the governing equation takes the form
\begin{equation}
\partial_t \psi =  -\alpha \left( \frac{\psi_{i+1} - \psi_{i-1}}{2 \Delta x} \right).
\end{equation}

It is worth noting that only the spatial components of the wave equation are discretized using the central difference scheme while the time evolution is advanced using a fixed time step $\Delta t = 0.05$.

We can use the same principle for the second model $E_2$ by extending the Taylor series expansion to include higher-order terms $\sin k \approx k - \frac{k^3}{3!}$ and $\cos k \approx 1 - \frac{k^2}{2}$ and using the binomial expansion to get
\begin{equation}
\label{eq:dispersion2_real_space}
E_2  = -i \hbar (1-V) \partial_x + i \hbar^3 \left(\frac{1}{8} + \frac{V}{6}\right)\partial^3_x.
\end{equation}

Substituting Eq.~(\ref{eq:dispersion2_real_space}) to the TDSE yields
\begin{equation}
i \partial_t \psi = -i (1-V(x)) \partial_x \psi + i  \left(\frac{1}{8} + \frac{V}{6}\right)\partial^3_x \psi,
\end{equation}
where the central difference approximation for the 3rd derivative is
\begin{equation}
\partial^3_x \psi \approx \frac{\psi_{i+2} - 2 \psi_{i+1} + 2 \psi_{i-1} - \psi_{i-2}} {2 \Delta x^3}.
\end{equation}

Thus, the full expression for the enhanced dispersion model $E_2$ is
\begin{equation}
\partial_t \psi =  -\alpha \left(\frac{\psi_{i+1} - \psi_{i-1}}{2 \Delta x} \right) + \beta  \left(\frac{\psi_{i+2} - 2\psi_{i+1 } + 2\psi_{i-1} - \psi_{i-2}}{2 \Delta x^3} \right),
\end{equation}
where $\beta= \left(\frac{1}{8} + \frac{V}{6} \right)$.

To advance the system in time, we employ the RK4 method:
\begin{equation}
\frac{d \psi}{dt} = F(\psi),
\end{equation}
where $F(\psi)$ represents the discrete spatial derivative operators for the wave equation derived from our respective dispersion relations ($E_1$ or $E_2$). The RK4 method approximates the solution by evaluating this function at four different points within each time step. The four intermediate slopes are computed as
\begin{align}
f_1 &= F(\psi^n), \\[6pt]
f_2 &= F\left(\psi^n + \tfrac{\Delta t}{2}f_1 \right), \\[6pt]
f_3 &= F\left(\psi^n + \tfrac{\Delta t}{2}f_2 \right), \\[6pt]
f_4 &= F\left(\psi^n + \Delta t\, f_3 \right),
\end{align}
where $f_1, f_2, f_3, f_4$ are the slope estimates at the beginning, midpoint (twice), and end of the time step, respectively. The wavefunction at the next time step $\Delta t$ is then computed as:
\begin{equation}
\psi^{n+1} = \psi^n + \frac{\Delta t}{6}\left(f_1 + 2f_2 + 2f_3 + f_4\right).
\end{equation}

\begin{figure}
    \centering
    \includegraphics[width=1.0\linewidth]{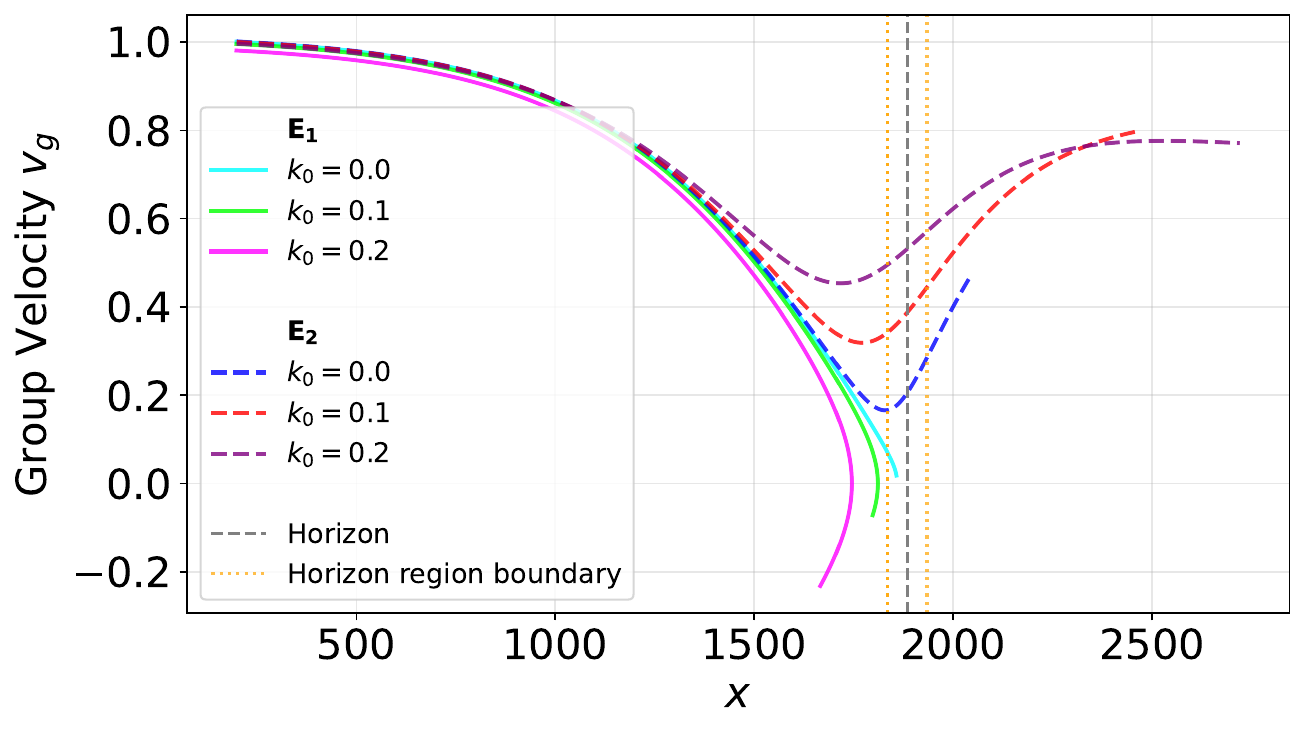}
    \caption{Group velocity $v_g$ versus position $x$ for $E_1$ and $E_2$ across initial momenta $k_0 = 0.0, 0.1, 0.2$. Orange dotted lines mark the horizon dwell zone boundaries ($x_h \pm 50$). $E_1$ shows velocity reversal ($v_g < 0$) indicating backward motion, with $k_0=0$ experiencing extreme deceleration near the horizon. $E_2$ maintains positive velocities throughout, with all packets slowing but continuing forward motion across the horizon region.}
    \label{fig:group_velocity}
\end{figure}

We implemented the dynamical evolution of the wave packet starting at the initial position $x_0 = 200$. Using the first model, the wave packet trajectories exhibit distinct behaviors depending on the initial momentum $k_0$. Figure~\ref{fig:wave_trajectories} shows that the wave packet with $k_0=0.0$ (cyan) approaches the horizon more closely at $x=1857.0$ compared to $k_0=0.1$ (lime) and $k_0=0.2$ (magenta) reaching its closest point at $x=1810.2$ and $x=1744.9$ respectively. Far from the horizon where $V(x) \approx 0$, the group velocity reduces to $v_g \approx \cos(k)$, and all packets propagate normally. As the wave packet move toward the horizon, $V(x)$ increases and the factor $\alpha=(1-V(x))$ causes the local $v_g$ to diminish continuously. At the horizon $V(x)=1$ which is the critical point where the dispersion collapses, the prefactor vanishes and that $E_1=0$ for all $k$ so the energy band flattens completely. This means that no continuous positive-energy propagation channel connects the left and right sides. Therefore, a wave packet approaching from the Type-I region cannot find a forward propagating state of the same energy beyond the horizon.

\begin{table}[h!]
\caption{\label{tab:summary_result}
  Final wave packet position $x$ and group velocity $v_g$ for different initial momenta $k_0$ under the two dispersion models $E_1$ and $E_2$.}
\setlength{\tabcolsep}{0pt}
\renewcommand{\arraystretch}{1.2} \extrarowheight0.2ex
\begin{tabular*}{\linewidth}{@{\extracolsep{\fill}}cccr}
  \hline \hline
  model & $k_0$ & Position $(x)$ & \multicolumn{1}{c}{$v_g$} \\
  \hline
  \multirow{3}{*}{$E_1$}
        & 0.0 & 1856.96 & 0.01 \\
        & 0.1 & 1797.12 & -0.07 \\
        & 0.2 & 1666.01 & -0.23 \\
  \hline
  \multirow{3}{*}{$E_2$}
        &  0.0 & 2039.20 & 0.46 \\
        & 0.1 & 2465.83 & 0.79 \\
        & 0.2 & 2719.69 & 0.76 \\
  \hline \hline
\end{tabular*}
\end{table}

At the final simulation time, the $k_0=0.0$ packet remains nearly frozen at its closest approach. In contrast, the finite $k_0$ packets exhibit negative group velocities indicating backward drift of their center of mass positions as illustrated in Fig.~\ref{fig:group_velocity}(a). The results of the wave packet dynamics are displayed in Table~\ref{tab:summary_result}.

The negative group velocities observed do not indicate classical backward propagation of the entire wave packet but rather reflect an asymmetric redistribution of probability density following wave packet disintegration. As each packet stops and begins to spread, its constituent momentum components separate spatially. Components with $k<k_0$ penetrates further toward the horizon while $k>k_0$ remains further back. As the packet disintegrates over time, probability density accumulates in the backward region through both the initial velocity distribution asymmetry and backscattering process. The center of mass being a weighted average, consequently drifts leftward even though the dominant probability remains near the stopping point.

Unlike the first model which exhibits complete reflection at the horizon, the second model $E_2$ demonstrates transmission through the horizon. In this case, the dispersion relation remains continuous across the horizon which allows finite energy states to exist throughout the spatial domain. Consequently, each wave packet crosses the horizon as illustrated in Fig.~\ref{fig:wave_trajectories}.

The transmission dynamics can be understood from an energy-based perspective. The zero-energy packet exhibits the strongest coupling to the horizon while finite-energy packets possess sufficient energy to transit the horizon region more rapidly. From an analog gravity perspective, this energy dependent behavior reflects the distinction between massless and massive particle dynamics near event horizons. The zero-energy packet represents the massless quasiparticles following null geodesics in the effective spacetime defined by $V(x)$. On the other hand, finite-energy packets behave more like massive particles with timelike geodesics experiencing weaker effective gravitational binding and crossing the horizon more readily.

A striking feature of the second model is the recovery of the $v_g$ after horizon crossing. Figure~\ref{fig:group_velocity}(b) shows that all transmitted wave packet experience deceleration approaching the horizon. However, beyond the horizon region, $v_g$ increases substantially as summarized in Table~\ref{tab:summary_result}.

In both our models, comparison reveals a consistent trend that follows the deviation between numerical and analytical group velocities. At $k_0=0.0$, the wave packet remains within the linear regime of the Weyl node following the geodesic-like propagation of massless quasiparticle near the horizon. As $k_0$ increases, higher order lattice effects becomes significant, producing systematic discrepancies between the numerical and analytical $v_g$. This behavior reflects a fundamental aspect of analog gravity in lattice systems that the correspondence to geodesic motion in a continuum spacetime holds only in the long wavelength limit. Our results confirm prior analyses where zero energy wave packets experiences significant slowdown near the horizon\cite{konye2022}. 

\subsection{WKB approximation}

As a complementary approach to the time-dependent RK4 simulations, we also examined the wave packet dynamics within the semiclassical WKB approximation. This method provides an analytical framework for understanding how spatial variations in the tilt profile $V(x)$ influence the propagation of quasiparticles in tilted Weyl semimetals.

\begin{figure}
    \centering
    \includegraphics[width=1.0\linewidth]{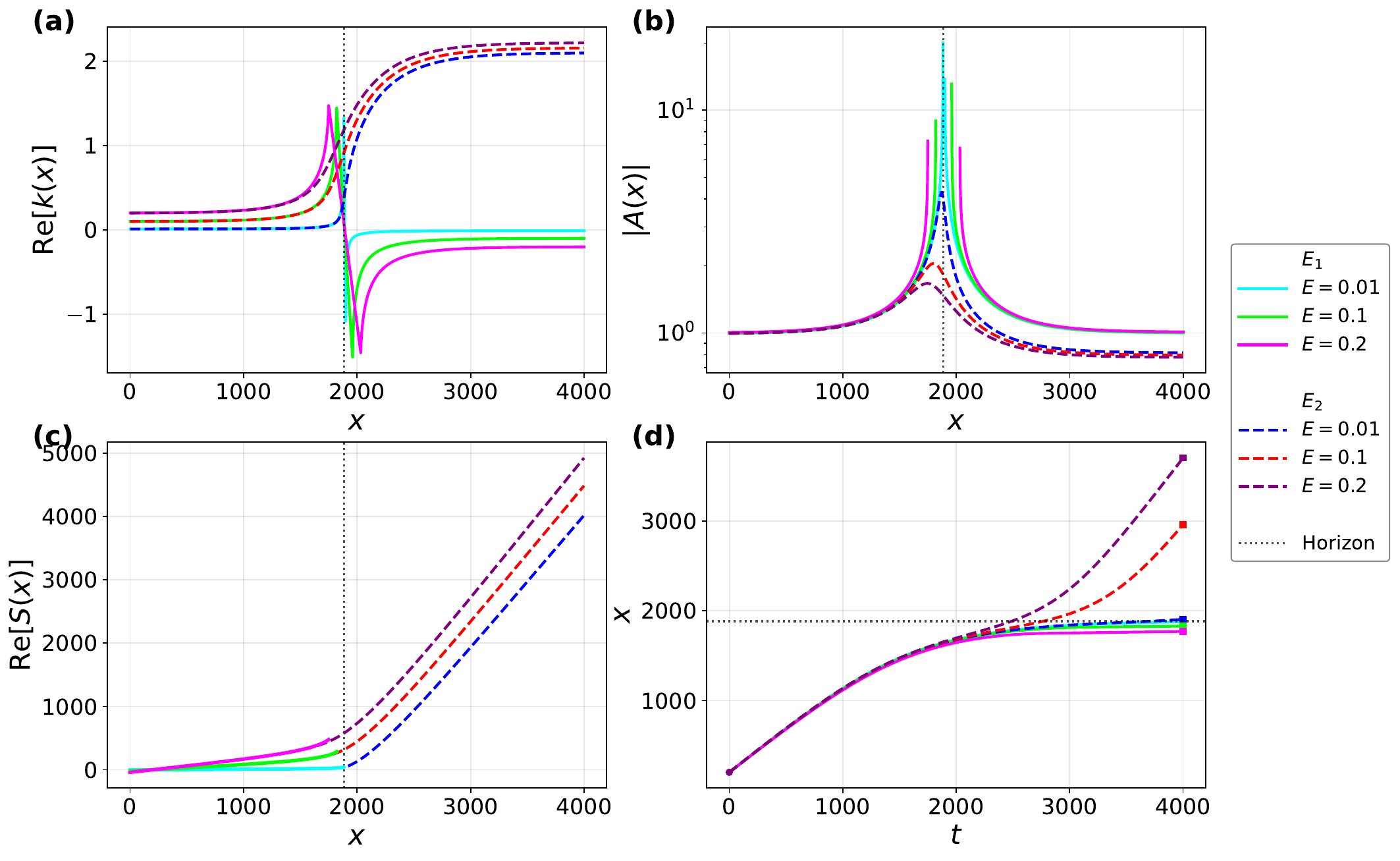}
    \caption{WKB analysis of horizon interactions using $E_1$ (straight lines) and $E_2$ (broken lines). (a) Local wavenumber, (b) amplitude divergence, (c) phase accumulation, and (d) trajectory dynamics for three energy values.}
    \label{fig:WKB}
\end{figure}

The WKB method is built upon the ansatz for the wavefunction,
\begin{equation}
\label{eq:WKB_ansatz}
\psi(x) = A(x) e^{i S(x)},
\end{equation}
where $A(x)$ is a slowly varying amplitude. The phase of the wavefunction, given by the classical action
\begin{equation}
\label{eq:phase}
S(x) = \int^{x}_{x_0} k(x') dx',
\end{equation}
is constructed via numerical integration outward from $x_0$ using the trapezoidal rule. The local wavenumber $k(x)$ satisfies the action $k(x)=\partial_x S$ and is determined by solving the Hamilton–Jacobi equation
\begin{equation}
E = H(x,k),
\end{equation}
where $H(x,k)$ is the classical Hamiltonian corresponding to the energy dispersion of the effective model. The group velocity, which dictates the wave packet's trajectory, can be solve by differentiating the energy dispersion
\begin{equation}
v_g = \partial_k E\bigg|_{k = k(x)}.
\end{equation}
In order to prevent numerical divergence, the magnitude of the group velocity is clamped to a small, positive minimum value near the horizon where $v_g \rightarrow 0$.

To determine the spatial variation of the amplitude $A(x)$, we make use of the probability current density. For any 1D wavefunction $\psi(x)$, the probability current is defined as
\begin{equation}
\label{eq:prob_current}
j(x) = \frac{\hbar}{m} \text{Im}\left(\psi^* \partial_x \psi \right).
\end{equation}
Solving for the spatial derivative gives us
\begin{equation}
\label{eq:spatial_derivative_ansatz}
\partial_x \psi = \left(\frac{dA}{dx} + \frac{i}{\hbar} A\frac{dS}{dx} \right)e^{iS(x)/ \hbar},
\end{equation}
where the complex conjugate of the wavefunction is
\begin{equation}
\label{eq:conjugate_ansatz}
\psi^* = Ae^{-iS(x)/ \hbar}.
\end{equation}
Plugging Eqs.~(\ref{eq:spatial_derivative_ansatz}) and (\ref{eq:conjugate_ansatz}) into Eq.~(\ref{eq:prob_current}) yields
\begin{equation}
j(x) = \frac{\hbar}{m} \text{Im}\left(Ae^{-iS(x)/ \hbar} \cdot \left(\frac{dA}{dx} + \frac{i}{\hbar}A\frac{dS}{dx} \right)e^{iS(x)/ \hbar} \right),
\end{equation}
where the exponential factors cancel out. This simplifies to
\begin{equation}
\label{eq:prob_current_2}
j(x) = \frac{\hbar}{m} \text{Im}\left(A \frac{dA}{dx} \right) + \frac{\hbar}{m} \text{Im}\left(\frac{i}{\hbar}A^2 \frac{dS}{dx} \right),
\end{equation}
where the first term is purely real because A(x) is real. The imaginary part of a real number is zero so that $\text{Im}\left(A \frac{dA}{dx} \right) = 0$. On the other hand, the second term has a factor of $i$ making it purely imaginary. The imaginary part of an imaginary number is itself, thus $\text{Im}\left(\frac{i}{\hbar} A^2 \frac{dS}{dx} \right) = \frac{1}{\hbar} A^2 \frac{dS}{dx}$. Thus, we can simplify Eq.~(\ref{eq:prob_current_2}) to
\begin{equation}
\label{eq:prob_current_3}
j(x) = \frac{1}{m} A^2 \frac{dS}{dx}.
\end{equation}

Recognizing that in the Hamilton-Jacobi formulation of classical mechanics, $\frac{dS}{dx}$ is the classical momentum $p$. In the semiclassical (WKB) limit, we treat the wave packet as a point-like particle. The velocity is the rate of change of its position, which is the group velocity. From classical mechanics, we also know that the particle velocity is $v=\frac{p}{m}$. Since we work on natural units where $\hbar = 1$, thus $p=k$ and that
\begin{equation}
v=\partial_p E = \partial_k E,
\end{equation}
where the right-hand side by definition is the group velocity $v_g$. Thus, the classical particle $v$ is numerically identical to the group velocity $v_g$ derived from the dispersion relation. Therefore, we make the identification
\begin{equation}
\frac{1}{m} \frac{dS}{dx} = \frac{p}{m} = v_g
\end{equation}

Substituting this into Eq.~(\ref{eq:prob_current_3}) and noting that the probability density $\rho(x) = |\psi(x)|^2 = [A(x)]^2$ gives us
\begin{equation}
j(x) = v_g|\psi|^2.
\end{equation}

For a stationary state, $|\psi|^2$ is time-independent so that $\partial_t \rho=0$ which makes the continuity equation 
\begin{equation}
\partial_t \rho + \partial_x j = 0
\end{equation}
reduce to $\partial_x j=0$. Since current is conserved,
\begin{equation}
j(x) = v_g |\psi(x)|^2 = \text{constant},
\end{equation}
which constrains the amplitude, yielding
\begin{equation}
A(x) = \frac{C}{\sqrt{|v_g|}},
\end{equation}
where $C$ is a normalization constant we will set to 1.

Thus, the WKB wavefunction of Eq.~(\ref{eq:WKB_ansatz}) takes the form of
\begin{equation}
\psi_{\mathrm{WKB}}(x) = \frac{C}{\sqrt{|v_g|}}\exp\left(-\frac{(x - x_0)^2}{2\sigma^2}\right)\exp\left(i \int_{x_0}^{x} k(x'), dx'\right),
\end{equation}
where the Gaussian envelope centered at $x_0$ with width $\sigma$ ensures spatial localization of the wave packet.

The two effective models differ only in their underlying dispersion relations that dictate the functional forms of $k(x)$ and $v_g$. Although both reduce to the same low-energy form as $k \rightarrow 0$, their global behaviors are qualitatively distinct. In the first model (Eq.~(\ref{eq:energy-disp})), the local wavenumber admits to a closed-form expression,
\begin{equation}
k(x) = \arcsin(E_0 / (1 - V(x))),
\end{equation}
while the second model (Eq.~(\ref{eq:energy-disp2})) requires numerical solution at each spatial point
\begin{equation}
f(k) = \sqrt{\sin^2 k + (1-\cos k)^2} - V(x)\sin k - E_0 = 0,
\end{equation}
where $E_0$ is the fixed energy eigenvalue of the wave packet. Using the Newton-Raphson method, the iterative scheme follows
\begin{equation}
k_{n+1} = k_n - \frac{f(k_n)}{f'(k_n)},
\end{equation}
where the derivative $f'(k) = \partial E_2/\partial k$ is
\begin{equation}
\label{WKB_vg}
f'(k) = \frac{\sin k}{\sqrt{\sin^2 k + (1-\cos k)^2}} - V(x)\cos k.
\end{equation}

Starting from the position $x_0$ where $V=0$, we obtain the initial wavenumber using $k_0=\text{min}(\text{max}(E,0.05),0.5)$ as a seed value. For subsequent spatial positions, we use the converged solution from the previous point as the initial guess where we exploit the smooth spatial variation of $V(x)$ to accelerate the convergence. The algorithm propagates both rightward and leftward from $x_0$.

The Newton-Raphson method succeeds in finding real-valued solutions at all spatial positions for all three energies including at and beyond the horizon. This contrasts sharply with the first model where solutions become undefined for $V \geq 1$.

Despite this added complexity, the general WKB structure where the dynamics is governed by the group velocity $v_g(x)$ and the integrated phase $S(x)$ remains intact. This will provide us with a direct means of comparing the two models.

We first investigate the simplest realization of an analog black hole horizon using the 1D tilted WSM described by the Hamiltonian in Eq.~(\ref{eq:Weyl_Hamiltonian_1}). We analyze the wave packet dynamics for three representative energies: $E_0=0.01$ (cyan), 0.1 (lime) and 0.2 (magenta) presented in Fig.~\ref{fig:WKB}.

In Fig.~\ref{fig:WKB}(d), we can see the classical trajectories of each packets moving with different speeds corresponding to their energies. As the packets approach their respective turning points, they slow down continuously, with their position curves flattening out. The turning point is determined by the condition $|E_0/(1-V)|=1$, which requires $V_{turn}\approx 1-E_0$. This makes the lower energy wave packet propagate exponentially closer to the horizon. This is because smaller energy allows the system to reach regions where $V$ is closer to 1 before the condition is violated. Eventually, all packets experience similar deceleration characteristics. This suggests that the dynamics are dominated by the spatial structure of the tilt rather than the initial energy of the wave packet.

As the tilt $V(x)$ increases towards the horizon, it demands that $k$ must increase to compensate for the vanishing $1-V$ term. This momentum gain is required to maintain the energy $E_0=(1-V)\sin(k)$ constant. Figure~\ref{fig:WKB}(a) shows this momentum increase explicitly.

Here we encounter a subtle but important point: while the lowest energy packet penetrates closest to the horizon, it reaches a lower final momentum than higher energy packets. The resolution lies in the tilt approaching unity and the momentum approaching the band edge. Lower energy packets are limited by how close they can get to $V=1$, while higher energy packets are limited by how close $k$ can get to $\pi/2$ where $\sin(k)$ saturates. The balance between these two constraints determines where each packet stops.

This momentum increase has a direct impact on the phase of the wave packet. Since wavelength is inversely proportional to momentum via de Broglie wavelength, this compresses the wave packet spatially. This extreme compression is the analog of gravitational blueshift for an infalling observe. For an external observer, this manifests as redshift of any signals emitted near the horizon. This wavelength compression is shown in Fig.~\ref{fig:WKB}(c). The slope of these curves equals the local wavenumber at each point manifested in Eq.~(\ref{eq:phase}). All three curves start with gentle slopes but then steepen as they approach the horizon.

This squeezing effect is governed by the conservation of probability flux where the current $j=|A|^2 v_g$ must remain constant. As the group velocity $v_g \rightarrow 0$ near the horizon, the amplitude $A(x)$ must increase to compensate. In particular, $A(x) \propto 1/\sqrt{v_g}$ implies that the amplitude diverges as $v_g$ approaches zero. This divergence behavior can be seen in Fig.~\ref{fig:WKB}(b).

The fundamental distinction between the first and second model emerges from the spectral properties at the horizon. The persistence of positive energy states at $V=1$ demonstrates that the spectrum does not collapse unlike the first model where $E_1(k,V=1)=0$ for all $k$. The dispersion term $\sqrt{\sin^2(k) + (1-\cos(k))^2}$ in Eq.~(\ref{eq:energy-disp2}) provides an energy contribution that remains finite at non-zero momenta even when the tilt-dependent term vanishes preventing spectral collapse. As mentioned earlier, as the tilt $V(x) \rightarrow 1$, the only way to maintain constant $E_0$ is for $k$ to increase. This forces $k$ into ranges where $\cos(k)<0$ which makes the tilt term contribute positively to $v_g$. Thus, the transmission occurs not despite the increasing tilt beyond the horizon but partly because of it.

This spectral continuity is reflected in the wavenumber evolution. Figure~\ref{fig:WKB}(a) shows $k(x)$ across the entire spatial domain including the classically forbidden region. Crucially, we observe that at every grid point, there exists real-valued $k(x)$ including $x>x_h$. The discontinuity across $x_h$ is $\Delta k = |k(x^-_h)-(x^+_h)|=0.017$, $0.008$ and $0.006$ for the three energies representing relative changes of $4.4\%$, $0.9\%$ and $0.5\%$, well within numerical tolerance. Here, $k(x^-_h)$ $\left(k(x^+_h)\right)$ represents the wavenumber at the grid point just before (after) the horizon.

Figure~\ref{fig:WKB}(b) displays the WKB amplitude $|A(x)|$ where all amplitude peaks occur where $v_g$ is minimum. The probability density concentrates where the wave packet moves slowest but unlike the first model, this pile-up is mild because $v_g$ remains finite.

As wave packets crossed the horizon, the amplitude decreases. At $x_h$ itself, amplitudes are continuous and do not exhibit divergent behavior. Beyond the horizon, amplitudes decreases further which is consistent with the velocity recovery in Fig.~\ref{fig:group_velocity}.

The classical action in Eq.~(\ref{eq:phase}) is computed via trapezoidal integration of the wavenumber field as shown in Fig.~\ref{fig:WKB}(c). The phase gradient $dS/dx=k(x)$ changes smoothly across $x_h$. This smooth transition measured directly from numerical integration, provides independent confirmation of wavenumber continuity. If $k(x)$ were discontinuous, the phase would exhibit a jump at $x_h$.

\begin{table}[h!]
\caption{\label{tab:transmission_result}
  Transmission parameters for the second model.}
\setlength{\tabcolsep}{0pt}
\renewcommand{\arraystretch}{1.2} \extrarowheight0.2ex
\begin{tabular*}{\linewidth}{@{\extracolsep{\fill}}ccccc}
  \hline \hline
  Energy & $t_{\mathrm{cross}}$ & $x_{\mathrm{f}}$ & $\Delta x_{\mathrm{trans}}$ & $v_{g,\mathrm{final}}$ \\
  \hline
  0.01 & 3721.1 & 1905.8 & 21.3  & 0.0964 \\
  0.1  & 2768.8 & 2960.0 & 1075.5 & 1.4741 \\
  0.2  & 2490.3 & 3705.1 & 1820.6 & 1.6422 \\
  \hline\hline
\end{tabular*}
\end{table}

Figure~\ref{fig:WKB}(d) presents the classical trajectories $x(t)$ obtained by solving $dx/dt=v_g(k(x),V(x))$. All three trajectories exhibit horizon crossing as summarized in Table~\ref{tab:transmission_result}. Here, $t_{cross}$ is the time at which $x(t)$ first exceeds the horizon, $x_f$ is the final position at simulation termination ($t=4000$) and $\Delta x_{trans}=x_f-x_h$ quantifies the transmission distance. Measured velocities immediately before and after horizon passage differ by less than $1\%$ for all energies. The packets $v_g$ decreases during the approach at $x_h$ from initial $1.0$ to minimum $0.054-0.358$. It then enters the horizon with finite velocity  ($0.062-0.460$) and subsequently accelerates beyond with final velocities ranging $0.096-1.642$. The higher energy packets cross earlier and penetrate further compared to lower energy packet. Final velocities also increase with energy, with the two higher energies achieving velocities exceeding their initial values.

\subsection{Scattering analysis}

To quantify the scattering properties at the horizon, we analyze the spatial distribution of the probability density of the wave packet $|\psi(x,t_f)|^2$ at the final simulation time $t_f=3500$ to ensure complete scattering dynamics have occurred.

\begin{figure}
    \centering
    \includegraphics[width=1.0\linewidth]{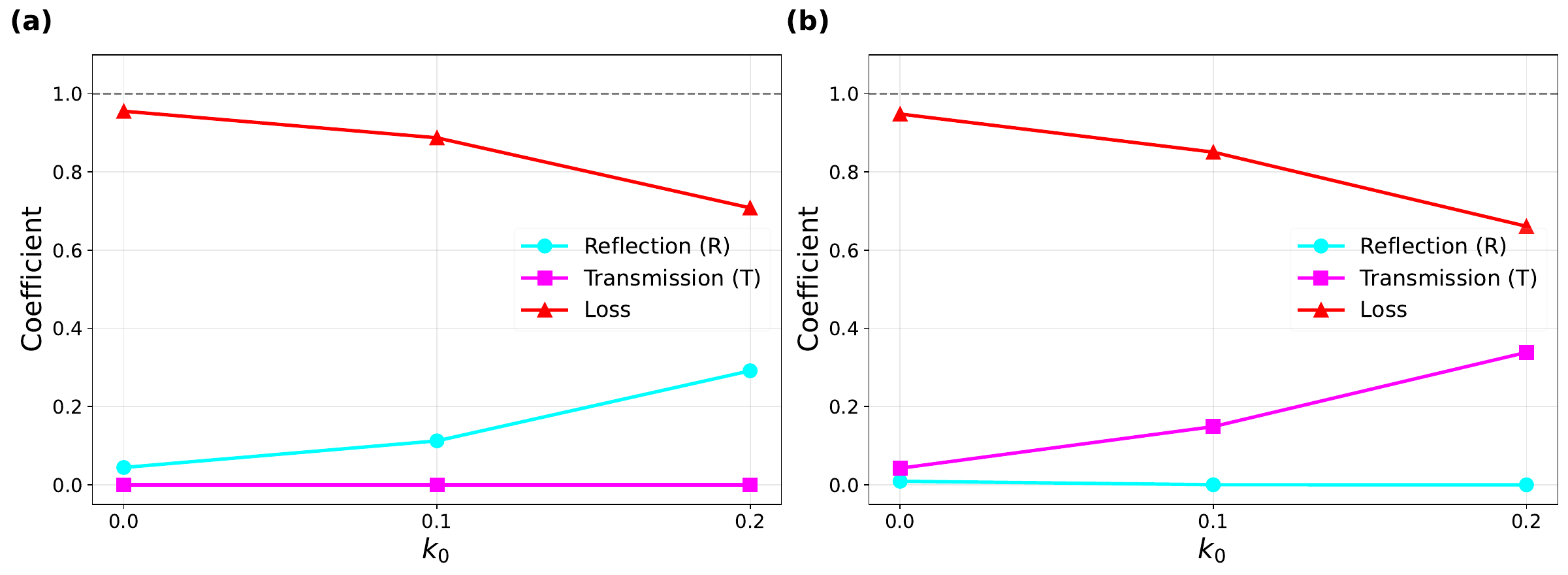}
    \caption{Scattering coefficients versus initial momentum $k_0$ for (a) $E_1$ and (b) $E_2$ models. First model exhibits complete reflection (R=1), while second model shows significant transmission (T>0).}
    \label{fig:combined_scattering}
\end{figure}

Due to the spatially varying tilt $V(x)$, the effective Hamiltonian is non-Hermitian. Standard quantum mechanics demands that the system's Hamiltonian be Hermitian. Conventional Hermitian quantum mechanics has been the foundation of quantum theory which to many of us, speaking of non-Hermitian Hamiltonian has no physical meaning. However, non-Hermitian Hamiltonian arise naturally as effective descriptions of open quantum systems~\cite{torres2019} including photonic \cite{ozawa2019, zhu2020} and Weyl systems \cite{yang2022, bergholtz2019} where gain and loss are inherently present.

As a result, probability is not conserved during wave packet evolution $\frac{dP}{dt} = \frac{d}{dt} \int |\psi(x,t)|^2 dx \neq 0$,
where $P(t)$ denotes the probability density at time $t$. This deviation from unitary serves as a quantitative measure of non-Hermitian dynamics induced by the position-dependent tilt. To preserve the physical effect of non-Hermiticity, no artificial normalization of $\psi$ is performed during the simulation. The total probability $P(t)$ is instead monitored directly, allowing the intrinsic loss dynamics to manifest naturally.

The system is partitioned into reflection ($x<x_h$) and transmission ($x>x_h$) regions relative to the horizon at $x_h \approx 1885$. For each momentum $k_0$, we compute the corresponding reflected and transmitted probabilities as
\begin{align}
    P_{\text{ref}} &= \int_{x<x_h} |\psi(x,t_f)|^2 dx, \\
    P_{\text{trans}} &= \int_{x>x_h} |\psi(x,t_f)|^2 dx.
\end{align}

\begin{table}[h!]
\caption{\label{tab:scattering_result}
Scattering coefficients for the $E_1$ and $E_2$ models. 
The reflection ($R$), transmission ($T$), and loss ($L$) coefficients satisfy $R + T + L = 1$.}
\setlength{\tabcolsep}{0pt}
\renewcommand{\arraystretch}{1.2} \extrarowheight0.2ex
\begin{tabular*}{\linewidth}{@{\extracolsep{\fill}}ccccc}
  \hline \hline
  Model
  & $k_0$ & $R$ & $T$ & $L$ \\
  \hline
  \multirow{3}{*}{1}
  & 0.0 & 0.0443 & 0.0000 & 0.9557 \\
  & 0.1 & 0.1124 & 0.0000 & 0.8876 \\
  & 0.2 & 0.2917 & 0.0000 & 0.7083 \\
  \hline
  \multirow{3}{*}{2}
  & 0.0 & 0.0091 & 0.0424 & 0.9485 \\
  & 0.1 & 0.0002 & 0.1489 & 0.8509 \\
  & 0.2 & 0.0000 & 0.3388 & 0.6612 \\
  \hline \hline 
\end{tabular*}
\end{table}

From these, the scattering coefficients are defined relative to the initial probability $P_{\text{init}}=1$ as
\begin{align}
    R &= \frac{P_{\text{ref}}}{P_{\text{init}}}, \\
    T &= \frac{P_{\text{trans}}}{P_{\text{init}}}.
\end{align}

Finally, probability loss is quantified by
\begin{equation}
    L=1-(R+T)=\frac{P_{\text{init}} - P(t_f)}{P_{\text{init}}},
\end{equation}
where $P(t_f)=P_{\text{ref}} + P_{\text{trans}}$ is the total remaining probability at $t_f$. Here, $L>0$ indicates net absorption of probability density during propagation. These quantities satisfy $R+T+L=1$, ensuring that the spatial integration correctly accounts for all probability. Table~\ref{tab:scattering_result} summarizes the scattering coefficients for both models across the three initial momenta.

The data reveal two striking features. First, the two models exhibit qualitatively opposite scattering behavior. The first model shows complete transmission suppression, while the second model permits forward transmission. Second, both models suffer substantial probability losses that dominate the dynamics of the surviving probability determining whether the wave packet reflects or transmits.

In the first model, transmission is completely suppressed for all values of $k_0$ as seen in Fig.~\ref{fig:combined_scattering}(a). The physical mechanism is transparent from the dispersion relation itself. When $V$ reaches its critical value, the energy vanishes identically for all momenta, eliminating propagating modes beyond $x_h$ and rendering forward transmission impossible.

The reflection coefficient vary strongly with momentum. This trend might initially suggest that reflection becomes more efficient at higher momenta but the correct interpretation is precisely the opposite. The reflection mechanism is identical across all $k_0$ values, but the incident wave reaching the horizon region is not the same for each $k_0$ which is the actual reason for the observed trend. 

As previously discussed at Sec.~\ref{sec:RK4}, the $k_0=0.0$ packet penetrates closest to the horizon region. Here, the dissipation rate $\partial_x V$ reaches its maximum value at $x=x_h$. This results in $95.6\%$ probability loss, leaving only $4.4\%$ to reflect. In contrast, higher momentum packets cease forward propagation further from the horizon due to earlier dispersive spreading. For instance, $k_0=0.2$ packet turns back well before entering the peak dissipation zone allowing $29.2\%$ to survive and reflect, not because the reflection process is more efficient, but rather from the reduced absorption due to their shallower penetration into the horizon.

Turning to the second model, we observe a qualitatively different scattering dynamics. Here, transmission is no longer forbidden and reflection is negligible as shown in Fig.~\ref{fig:combined_scattering}(b).

The second model exhibits finite transmission up to $34\%$ with reflection suppressed below detection limits for higher $k_0$ packets. Even the wave packet with lowest $k_0$ which barely penetrates the horizon manages to transmit $4.2\%$ of its probability while reflecting less than $1\%$. This behavior is not a quantitative shift but a qualitative transformation of the horizon from an impenetrable wall to a partially transmitting membrane. This near-total suppression of backscattering combined with finite transmission indicates that the horizon using the second model is not scattering in the traditional sense but is simply allowing forward propagation with attenuation. Wave packets encountering the horizon do not bounce back but passes through with probability $T$ or dissipates with probability $L$. The backward propagation has been essentially closed.

Despite this transmission behavior, substantial probability loss also persists in the second model. Examining the two models, we find them comparable with $66-95\%$ for the second model compared to $71-96\%$ for the first model. This immediately tells us that dissipation is governed primarily by the tilt rather than the specific dispersion relation.

\subsection{Dwell time calculation}

In addition to scattering probabilities, we compute the dwell time $\tau$ for which the wave packet remained in the vicinity of the horizon during its evolution. The instantaneous position of the wave packet is characterized by its center of mass $\langle x \rangle (t)$ which represents the expectation value of position at time $t$.

\begin{figure}
    \centering
    \includegraphics[width=1.0\linewidth]{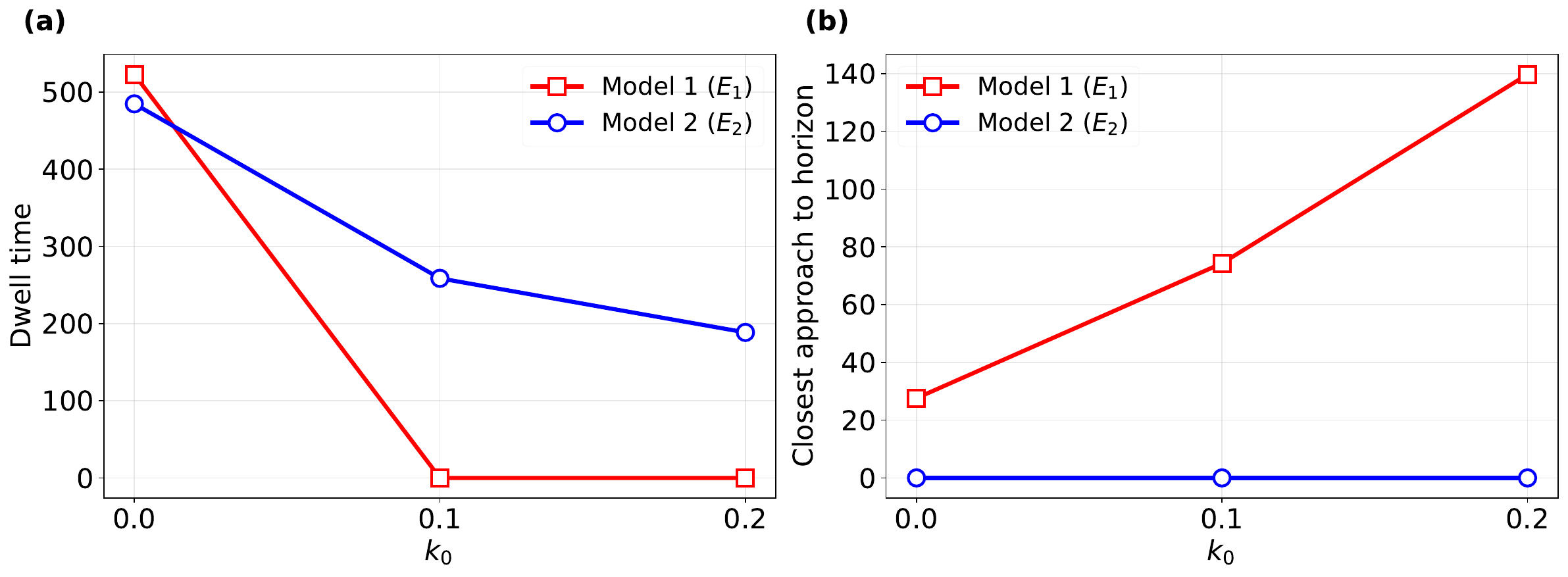}
    \caption{Comparison of (a) dwell time and (b) closest approach to the horizon for different initial momenta $k_0$ under the two dispersion models $E_1$ and $E_2$.}
    \label{fig:dwell}
\end{figure}

The dwell region is defined as a spatial window centered at the horizon
\begin{equation}
    D={x \mid|x-x_h|<\delta}
\end{equation}
where $\delta=50$ is the dwell threshold for the horizon region. At each time step, we track whether the wave packet center lies within the dwell region. The total dwell time is obtained as
\begin{equation}
    \tau=\sum_j\left(t_{out}^{(j)} - t_{in}^{(j)} \right),
\end{equation}
summing over all dwelling interactions. If the packet remains within $D$ at the final simulation time, the last interval is closed at $t=t_f$.

\begin{table}[h!]
\caption{\label{tab:dwell_comparison}
  Dwell time $\tau$ and minimum distance to the horizon $x_{\min}$ for different initial momenta $k_0$ under the two dispersion models $E_1$ and $E_2$.}
\setlength{\tabcolsep}{0pt}
\renewcommand{\arraystretch}{1.2} \extrarowheight0.2ex
\begin{tabular*}{\linewidth}{@{\extracolsep{\fill}}ccccc}
  \hline \hline
  Model
  & $k_0$ & $\tau$ & Entries & $x_\mathrm{min}$ \\
  \hline
  \multirow{3}{*}{$E_1$}
  & 0.0 & 522.8 & 1 & 27.6 \\
  & 0.1 & 0.0   & 0 & 74.3 \\
  & 0.2 & 0.0   & 0 & 139.7 \\
  \hline
  \multirow{3}{*}{$E_2$}
  & 0.0 & 484.7 & 1 & 0.0 \\
  & 0.1 & 258.6 & 1 & 0.0 \\
  & 0.2 & 188.5 & 1 & 0.0 \\
  \hline \hline
  \end{tabular*}
\end{table}

The dwell time measurements reveal the temporal extent of wave packet interaction with the horizon region. Table~\ref{tab:dwell_comparison} summarizes the dwell time statistics and closest approach distances for both models across all initial momenta. The results reveal different horizon interaction patterns between the two dispersion models, reflecting their contrasting spectral properties and transmission characteristics established in previous sections.

In the case of the first model, only the $k_0=0.0$ packet enters the horizon region spending $\tau=522.8$ (see Fig.~\ref{fig:dwell}(a)) time units within $\delta=50$ of the horizon dwell region and approaching to within $x_{min}=27.6$ (see Fig.~\ref{fig:dwell}(b)) lattice units of the horizon. This extended dwelling represents the "freezing" behavior analogous to characteristic of particles near even horizons. 

On the other hand, the $k_0=0.1$ and $k_0=0.2$ packets never enter the dwell region. Their closest approaches are $x_{min}=74.3$ and $x_{min}=139.7$ units respectively which occurs well outside the $\delta=50$ threshold. This behavior directly reflects the dispersive spreading and momentum redistribution discussed in Sec.~\ref{sec:RK4}. The packet disintegrates before reaching the region of strong horizon interaction, with trailing components dominating the center of mass position and preventing further advance.

This correlation between dwell time and closest approach is evident. Only the $k_0=0.0$ packet which maintains coherence and propagates deeply, experiences significant dwelling. The inverse relationship between momentum and penetration depth establishes $k_0=0.0$ as the unique case exhibiting prolonged horizon interaction in first model.

In contrast, the second model presents a qualitatively different scenario where all packets enter and cross the dwell region with $x_{min}=0$ recorded for all $k_0$ values. The zero minimum distance indicates that the wave packet centers traverse through the horizon position $x=x_h$ during its evolution. 

However, the dwell times vary systematically with momentum. Dwell time $\tau$ decreases from 484.7 at $k_0=0.0$ to 258.6 at $k_0 = 0.1$, and further to 188.5 at $k_0 = 0.2$. The $k_0=0.0$ packet, experiencing maximum deceleration at the horizon takes the longest to traverse the 100 site-wide dwell window. The higher-energy $k_0 = 0.2$ packet completes the same spatial transit faster.

Comparing with the first model, the $k_0=0.0$ dwell times are remarkably similar, 522.8 for $E_1$ model versus 484.7 for $E_2$ model. This near-equivalence suggests that $k_0=0.0$ packets in both models experience comparable horizon interaction strengths during their initial approach and extended dwelling phase. The critical difference emerges in the outcome is that the first model forces eventual reflection after the dwelling period, while second model's continuous spectrum allows continued forward motion.

\section{Conclusion and Discussion}

We have investigated wave packet dynamics in 1D tilted Weyl semimetals through two distinct models that realize analog black hole horizons using complementary semi-classical WKB analysis and RK4 simulations. The first model $E_1$ exhibits complete spectral collapse at the horizon ($V=1$) where $E_1(k)=0$ for all momenta creating an impenetrable barrier with zero transmission \cite{konye2022}. Critically, only the $k_0=0.0$ packet approaches the analog horizon, penetrating to within 27.6 lattice units in the horizon region. Finite momentum packets undergo severe dispersive spreading, disintegrating before reaching the horizon region. Their apparent reflection represents decoherence in an inhomogeneous potential rather than gravitational scattering. The WKB analysis reveals extreme wavelength compression, amplitude divergence and vanishing group velocity at the horizon. 

On the other hand, second model maintains spectral continuity by preserving finite density of states at non-zero momenta even when $V=1$ \cite{konye2022}. All wave packets successfully cross the horizon with transmission coefficients of $4-34\%$ that increase systematically with $k_0$. The wavenumber, amplitude, and phase evolve continuously through the horizon, with group velocities remaining finite and recovering after crossing as energy conservation drives momentum growth beyond $k=\pi/2$ where $\cos(k)$ changes sign, converting the tilt from decelerating to accelerating force.

The scattering analysis reveals that the first model exhibits reflection-dominated behavior while second model shows transmission-dominated dynamics. Both models exhibit substantial probability loss $(66-96\%)$ arising from the non-Hermitian character of the position-dependent tilt representing dissipation into degrees of freedom beyond our 1D effective description. The near-identical loss fractions between models demonstrate that dissipation is controlled primarily by the spatial tilt gradient and exposure time rather than the spectral structure. Dwell time analysis provides temporal signatures complementing the scattering coefficients. First model's lowest $k_0$ packet dwells 523 time units near the horizon while higher $k_0$ packets never enter the horizon vicinity, reflecting early with zero dwell time. Second model exhibits slowed but continuous transit for all $k_0$, with dwell times of 189-485 decreasing monotonically with momentum. The near-identical dwell times for $k_0=0.0$ in both models despite opposite outcomes demonstrates that both systems create comparable "gravitational potentials" with similar interaction strengths differing only in whether the spectral structure permits reflection or transmission.

In summary, our work demonstrates that tilted Weyl semimetals can reproduce key aspects of black hole horizon physics. By tuning the Hamiltonian, one can access regimes spanning from impenetrable barriers to partially transmitting membranes. Future work should include investigation of more complex horizon profiles, multi-dimensional geometries, and connections to Hawking radiation and information paradox scenarios.

\begin{acknowledgments}
This work is supported by the Science and Technology Regional Alliance of Universities for National Development (STRAND) scholarship granted to M.A. Lozande by the Department of Science and Technology-Science Education Institute (DOST-SEI).

EAF gratefully acknowledges support from the Mindanao State University - Office of the Vice Chancellor for Research, Extension and Development through the Mamitua Saber Institute of Research and Creation under the Special Order No. 351 - OP Series of 2025. 
\end{acknowledgments}

\section*{Data Availability}

The Python codes used in this work are available in the Zenodo repository (https://doi.org/10.5281/zenodo.17727711)\cite{Lozande2025}.

\end{document}